\documentclass[11pt]{article}
\pdfoutput=1
\usepackage{graphicx}
\usepackage{epstopdf}
\usepackage{natbib}
\usepackage{amsmath}

\begin{document}

\title{Motion of a helical vortex}

\author{Oscar Velasco Fuentes \\
              Departamento de Oceanograf\'{\i}a F\'{\i}sica, CICESE, Ensenada, M\'exico\\
              Email address: ovelasco@cicese.mx}
              
\date{\small{Submitted to the European Journal of Mechanics B/fluids, 30 July 2015}}              
\maketitle

\begin{abstract}
We study the motion of a single helical vortex in an unbounded, inviscid, incompressible fluid. 
The vortex is an infinite tube whose centerline is a helix
and whose  cross section is a circle of small radius (compared to the radius of curvature) 
where the vorticity is uniform and parallel to the centerline.
Ever since Joukowsky (1912) deduced that this vortex translates and rotates steadily without change of form, 
numerous attempts have been made to compute these self-induced velocities. 
Here we use  Hardin's (1982) solution for the velocity field 
to find new expressions for the vortex's linear and angular velocities.
Our results, verified by numerically computing the Helmholtz integral and the Rosenhead-Moore approximation
to the Biot-Savart law, are more accurate than previous results over the whole range of values of
the vortex pitch and cross-section.
We then use the new formulas to study the advection of passive particles near the vortex;
we find that the vortex's motion and capacity to transport fluid depend on its pitch and cross section as follows:
a thin vortex of small pitch moves fast and carries a small amount of fluid; 
a thick vortex of small pitch moves at intermediate velocities and is a moderate carrier itself but it pushes fluid forward along the helix axis; and
a vortex of large pitch, whether thin or thick, moves slowly and carries a large amount of fluid.
\end{abstract}

%
%
%
\begin{figure}
\centering
\includegraphics[width=0.25\textwidth]{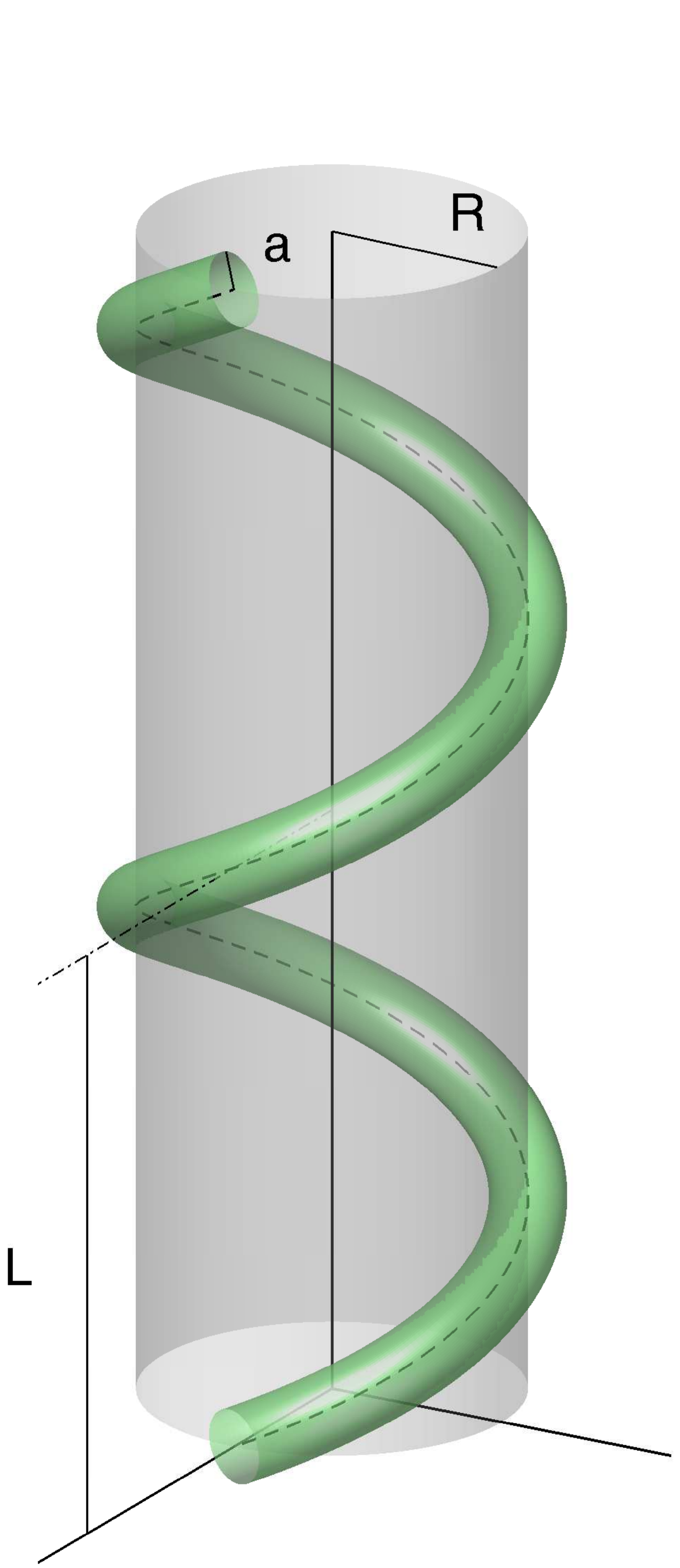}
\caption{
A segment of a thin helical vortex.
The vortex extends indefinitely in both directions and
its centerline is a helix of pitch $L$ and radius $R$ lying 
on the surface of an imaginary supporting cylinder.
}
\label{f:helix}   
\end{figure}

\section{Introduction}
\label{s:intro}

The realisation that concentrated helical vortices are frequent
features in natural and man-made flows has grown steadily
since \cite{Parsons} discovered that ``a small spiral vortex existed just behind 
the tips of the propeller blades'' used to generate thrust to propel a vessel.
The theoretical study of helical vortices has an even longer history.
Here we highlight the earliest results, but
see \cite{Ricca} for a detailed review of classical studies of 
the self-induced velocity of a single helical vortex in an inviscid, unbounded fluid (figure~\ref{f:helix}). 
\cite{Kelvin} found that one mode of vibration of a cylindrical vortex
is a wave that deforms the vortex axis into a helix of small radius and large pitch, and that this wave propagates along 
the axis with constant speed.
\cite{Fitzgerald} assumed that this result was valid for helices of any pitch 
and radius and speculated about the flow induced by the vortex: 
``there will be, on the whole, a flow along the inside of the spiral, but the motion 
of the fluid is complex.''
\cite{Joukowsky} showed that a helical vortex moves steadily without change of form
and found that its self-induced velocity is approximately equal to the velocity of an osculating ring vortex;
he further stated that any number of equal helical vortices symmetrically arranged with respect to a common axis
form a steadily moving arrangement.

\begin{figure}
\centering
\includegraphics[width=0.35\textwidth]{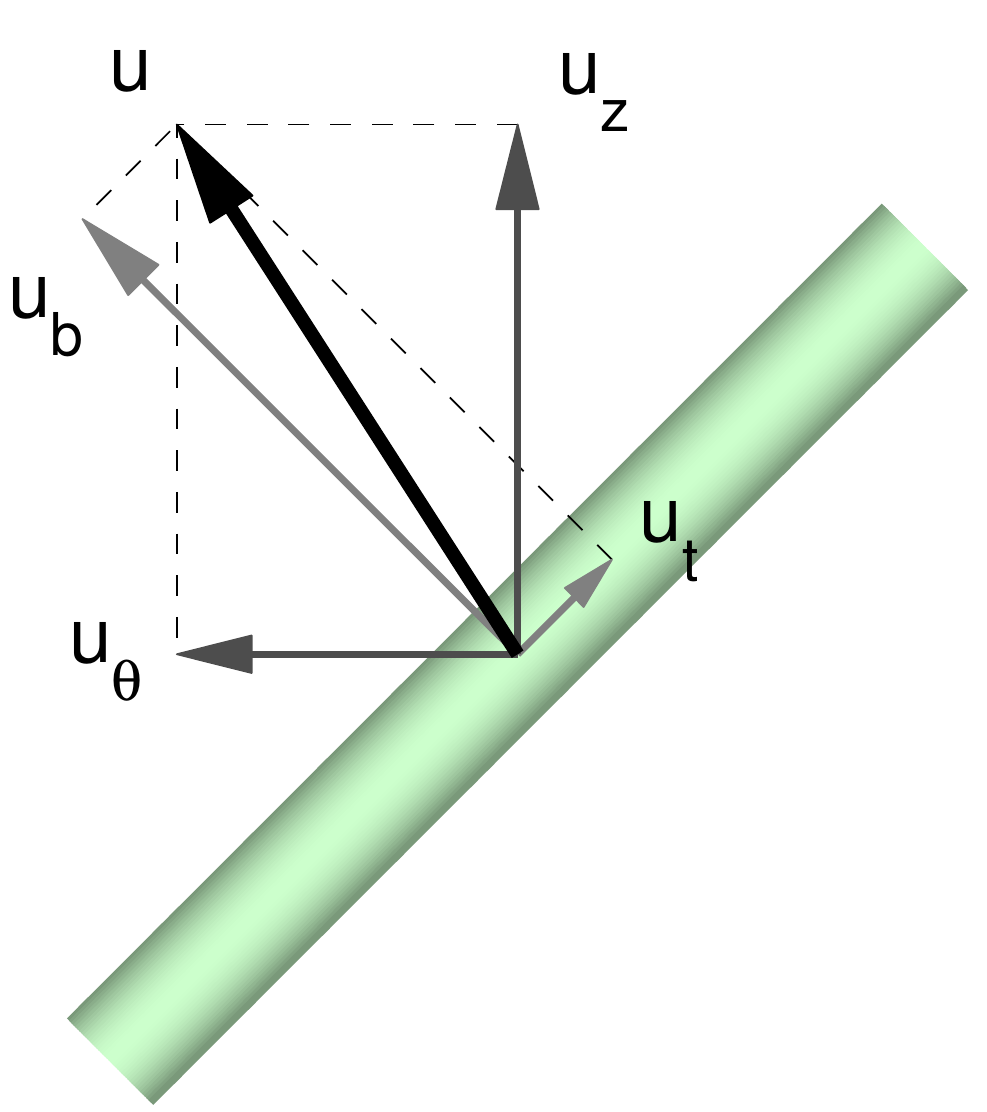}
\caption{
The self-induced velocity of a helical vortex, $u$, represented as the sum of either 
the azimuthal and axial components ($u_\theta$ and $u_z$, respectively)
or  
the tangential and binormal components ($u_t$ and $u_b$, respectively).
}
\label{f:systems}   
\end{figure}

In the last two decades there has been a renewed interest on helical vortices \citep{Ricca,Mezic,Kuibin,Boersma,Wood,Okulov}.
Some of these works have concentrated on the binormal component of the self-induced velocity and have implicitly or 
explicitly dismissed the tangential component (see figure~\ref{f:systems}).
This is unfortunate because accurate values of the full self-induced velocity
are essential for a proper analysis of the vortex stability  \cite[see, e.g.,][]{Widnall} or the
particle motion in the vicinity of the vortex \cite[see, e.g.,][]{Mezic,Andersen}.

The objectives of this paper are, first, to obtain expressions for the
vortex's linear and angular velocities ($U$ and $\Omega$, respectively) 
that are accurate over the whole range of pitch values
for vortices of small cross section (when compared to the radius of curvature of their centerline);
and, second, to apply the new formulas in the study of particle motion in the flow induced by helical vortices.
Section~\ref{s:motion} contains 
our theoretical calculations, which are based on the velocity field obtained by \cite{Hardin};
the verification of the new formulas for $U$ and $\Omega$ by numerical computation of the Helmholtz integral and the Rosenhead-Moore approximation to the Biot-Savart law;
and a comparison with previous results \citep{Joukowsky,DaRios16,Levy,Widnall,Mezic,Okulov}.
In section~\ref{s:topology} we determine the vortex's capacity to carry fluid by analysing
the helical stream function of \cite{Hardin} in a system that moves with the vortex.
We present our conclusions in section~\ref{s:conclusions}.

%
%

%
%
%

\section{Self-induced motion of the vortex}
\label{s:motion}

A helical vortex is a thin tube of infinite length whose centerline is a helix of uniform pitch lying on the surface of 
an imaginary supporting cylinder (see figure~\ref{f:helix}). 
The centerline of the vortex is given, in Cartesian coordinates, as follows:
\begin{eqnarray*}
x&=& R \cos \theta, \\
y&=& R \sin \theta, \\ 
z&=&  L\theta/2\pi, 
\end{eqnarray*}
where $\theta$ is the angle around the cylinder's axis,
$R$ is the radius of the helix
and $L$ is the pitch of the helix.

On the vortex's circular cross-section the vorticity is uniform in magnitude and direction, 
which is parallel to the centerline's tangent. Note that both, the circular shape and the
uniform vorticity, are leading-order approximations only: in a steady solution
of the Euler equations the vorticity varies linearly with the distance to the centre 
of curvature and the cross-section is a small perturbation of the circular shape.
The vortex thus has circulation $\Gamma=\pi a^2 \omega$, 
where $a$ is the radius of the cross-section and $\omega$ is the magnitude of the vorticity. 
The four parameters that uniquely define the vortex are then $\Gamma$, $R$, $L$ and $a$ but, 
since the value of $\Gamma$ merely changes the time scale of the evolution, 
we are left with two non-dimensional parameters: 
the vortex radius $\alpha=a/R$ and 
the vortex pitch $\tau=L/2\pi R$.

\subsection{Analytical results}
\label{s:analytical}

Because of the geometry of the problem the self-induced velocity of the vortex can be expressed in either 
cylindrical components, $u_\theta$, $u_z$, pointing in the azimuthal and axial directions, respectively;
or in natural components, $u_t$, $u_b$, pointing in the tangent and binormal directions, respectively 
(see figure~\ref{f:systems}).
We will use cylindrical coordinates instead of the now more popular natural coordinates
because the translation speed of the vortex, $U$, equals the axial component of the self-induced velocity, $u_z$,
whereas the angular velocity of the vortex, $\Omega$, equals the azimuthal component divided 
by the helix radius, $u_{\theta}/R$.

\cite{Hardin} found the velocity field produced by an infinitely-thin helical vortex.
His solution, expressed in cylindrical coordinates, is divided in 
an interior field (valid inside the supporting cylinder) and an exterior field (valid outside the supporting cylinder). Here we reproduce only
the azimuthal and axial components needed for the calculation of the vortex motion:
\begin{equation}
\label{e:utheta}
u_\theta(r,\phi)= 
\left \{
\begin{array}{l l} 
\dfrac{\Gamma R}{\pi r l} S_1(r,\phi) & \text{if } r<R \\
 & \\
\dfrac{\Gamma }{2 \pi l} + \dfrac{\Gamma R}{\pi r l} S_2(r,\phi) & \text{if } r>R 
\end{array}
\right.
\end{equation}
\begin{equation}
\label{e:uz}
u_z(r,\phi)= 
\left \{
\begin{array}{l l} 
\dfrac{\Gamma }{2 \pi l} - \dfrac{\Gamma R}{\pi l^2} S_1(r,\phi) & \text{if } r<R \\
 & \\
- \dfrac{\Gamma R}{\pi l^2} S_2(r,\phi) & \text{if } r>R 
\end{array}
\right.
\end{equation}
where $\phi=\theta-z/l$, $l=L/2\pi$ and
\begin{eqnarray*}
S_1(r,\phi) &=& \sum_{m=1}^{\infty} m K'_m \left( \frac{m R}{l} \right) I_m \left( \frac{m r}{l}\right) \cos m\phi \\
S_2(r,\phi) &=& \sum_{m=1}^{\infty} m K_m \left( \frac{m r}{l} \right)  I'_m \left( \frac{m R}{l}\right) \cos m\phi 
\end{eqnarray*}
are Kapteyn series involving the modified Bessel functions $K_m$ and $I_m$, and their
corresponding derivatives $K'_m$ and $I'_m$.

Since we are dealing with vortices of finite cross section, it is possible to compute the vortex's self-induced velocity 
by evaluating the velocity field at two diametrically-opposed points on the surface of the tubular vortex.
The uniformity of the vorticity on the cross section guarantees that the average of these velocities is  the actual 
velocity at the centerline, i.e. the velocity of the vortex. 
This approach has been previously used by \cite{Ricca}, \cite{Boersma}, and \cite{Okulov} for the computation 
of the binormal component; here we will use it for the computation of the axial and azimuthal components.
For simplicity, we choose to evaluate the velocity at points $(r,\theta,z)=(R \pm a, 0, 0)$ or, in helical coordinates, $(r,\phi)=(R \pm a, 0)$. Therefore, the self-induced motion
of the vortex is given by
\begin{eqnarray}
U           &=&\frac{1}{2 } \left[ u_z (R-a,0)+u_z(R+a,0) \right]  \label{e:U}\\
\Omega &=&\frac{1}{2 R} \left[ u_\theta(R-a,0)+u_\theta(R+a,0) \right]  \label{e:Om}
\end{eqnarray}

A technical issue with this method is that the series $S_1$ and $S_2$ converge slowly, particularly when approaching the vortex itself,
which hinders the computation of $U$ and $\Omega$. This problem was solved by \cite{Boersma}, who eliminated the singularities from $S_1$ and $S_2$
for the two particular points needed in the computation of the self induced motion:
\begin{eqnarray*}
S_1(R-a,0) &=&  \frac{1}{4}\frac{\tau^2}{(1+\tau^2)^{3/2}}\left(  - \frac{2}{\epsilon}+\ln(\epsilon)+\ln \left( \frac{\sqrt{1+\tau^2}}{2}   \right)\right)+\frac{\tau}{2}-\frac{\tau^2}{4} W(\tau) + o(1)\\
S_2(R+a,0) &=& \frac{1}{4}\frac{\tau^2}{(1+\tau^2)^{3/2}}\left(   \frac{2}{\epsilon}+\ln(\epsilon)+\ln \left( \frac{\sqrt{1+\tau^2}}{2}  \right)\right)-\frac{\tau^2}{4} W(\tau) + o(1) 
 \end{eqnarray*}
where $\epsilon=a/R(1+\tau^2)$, $o(1)$ represents an expression that tends to $0$ when $\epsilon \to 0$, and
\begin{displaymath}
W(\tau)= \int_0^\infty   \left[      \frac{\sin^2 t} {(\tau^2 t^2+\sin^2 t)^{3/2}} - \frac{1}{(1+\tau^2)^{3/2}} \frac{H(1/2-t)}{t}\right] dt
\end{displaymath}
with $H(\cdot)$ the unit step function. \cite{Boersma} computed $W(\tau)$ by numerical quadrature for 21 values of $\tau$ in the range $[0.01-10]$
and obtained approximate analytical forms for large values of $\tau$ (giving less than 1\% error for $\tau \ge 3$)
and for small values of $\tau$ (giving less than 1\% error for $\tau \le 0.4$).
Here we computed the value of $W(\tau)$ using the numerical method described by \cite{Boersma}.

%
%
%
\begin{figure}
\centering
\includegraphics[width=0.9\textwidth]{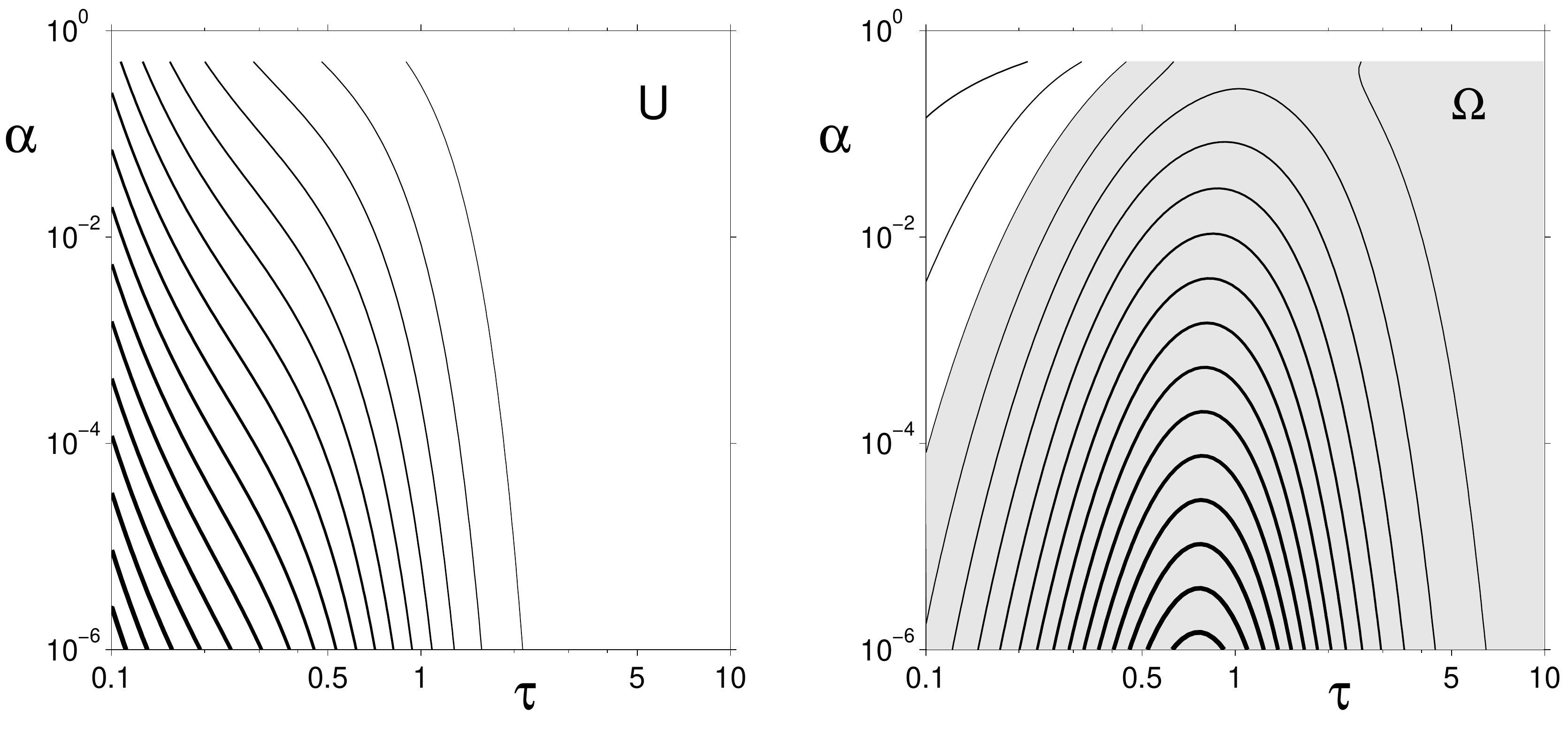}
\caption{
The linear and angular velocities of a helical vortex ($U$ and $\Omega$, respectively)
as functions of its pitch and radius ($\tau$ and $\alpha$, respectively).
The thick contours indicate higher absolute values of $U$ or $\Omega$; the grey areas indicate
the region of the parameter space where $\Omega$ is negative (clockwise rotation of the helix).
}
\label{f:OmUplane}   
\end{figure}

The substitution of these forms of $S_1$ and $S_2$ in (\ref{e:utheta})-(\ref{e:uz}) and 
the subsequent substitution of the resulting expressions for $u_\theta$ and $u_z$ in (\ref{e:U})-(\ref{e:Om}) 
give, after some algebra, the linear and angular velocities in nondimensional form:
\begin{eqnarray*}
U^*           &=&   \frac{    1}{(1+\tau^2)^{3/2}}           \left( \ln(2/\epsilon) - \ln(\sqrt{1+\tau^2}) + (1+\tau^2)^{3/2} W(\tau)  \right) \\
\Omega^* &=&   \frac{\tau}{(1+\tau^2)^{3/2}} \left(\frac{  2(1+\tau^2) +  \ln(2/\epsilon) - \ln(\sqrt{1+\tau^2}) -  (1+\tau^2)^{3/2} (2/\tau- W(\tau))  }{ \epsilon^2(1+\tau^2)^2-1  }\right)\\
\end{eqnarray*}
In dimensional form the self-induced motion of the vortex is given by
\begin{eqnarray}
U           &=&  \frac{\Gamma}{4 \pi R}     U^*   \label{e:Udim}\\
\Omega &=&  \frac{\Gamma}{4 \pi R^2}  \Omega^* \label{e:Omdim}
\end{eqnarray}

Figure~\ref{f:OmUplane} shows $U$ and $\Omega$ in the region $10^{-6}<\alpha<0.4$ and $0.1<\tau<10$. 
All vortices translate in the direction of the axial component of the vorticity
with a velocity $U$ that increases as their pitch $\tau$ and radius $\alpha$ decrease.
The vortices generally rotate in a direction opposite to the azimuthal component of the vorticity,
i.e. clockwise when seen from the direction in which the vortex translates. 
They rotate with an angular velocity $\Omega$ that, for a fixed radius $\alpha$, 
has a maximum absolute value when their pitch $\tau$ is about one and that, for a fixed $\tau$, 
increases in magnitude as $\alpha$ decreases.
Only relatively thick vortices of small pitch rotate in anti-clockwise sense
(upper left corner in figure~\ref{f:OmUplane}).

Note that the translation speed $U$ behaves as the helix's curvature, {\mbox{$1/R(1+\tau^2)$}},
which has a maximum value ($1/R$) when $\tau=0$
and then decreases to zero as $\tau$ goes to infinity.
Similarly, the magnitude of the rotation speed $\Omega$ behaves as the helix's torsion, {\mbox{$\tau/R(1+\tau^2)$},
which grows from zero when $\tau=0$ to its maximum value ($1/2R$) when $\tau=1$ and then 
decreases to zero as $\tau$ goes to infinity.

\subsection{Comparison with numerical results}
\label{s:numerical}

We verified equations (\ref{e:Udim}) and (\ref{e:Omdim}) by computing the
self-induced velocities of the vortex by numerical integration of the Helmholtz formula,
\begin{equation}
\label{e:Helmholtz}
\boldsymbol{u}(\boldsymbol {x}) = -\dfrac{1}{4 \pi}  \int
                     \dfrac{[\boldsymbol {x}-\boldsymbol {x'}] \times \boldsymbol {\omega} }
                         {   |\boldsymbol {x}-\boldsymbol {x'}|^2 } dV,
\end{equation}
and the Rosenhead-Moore approximation to the Biot-Savart law,
\begin{equation}
\label{e:RM}
\boldsymbol{u}(\boldsymbol {x}) = -\dfrac{\Gamma}{4 \pi}  \int 
                     \dfrac{[\boldsymbol {x}-\boldsymbol {r}(s)] \times d\boldsymbol {s} }
                         {\left( |\boldsymbol {x}-\boldsymbol {r}(s)|^2+\mu^2a^2 \right) ^{3/2}},
\end{equation}
where $a$ is the radius of the cross section of the tubular vortex and $\mu$ 
is a parameter that depends on the vortex local structure, i.e. its vorticity distribution, curvature and torsion.
Since, by our definition, a helical vortex has everywhere the same local structure $\mu$ is a constant but 
its value must  be determined.  
This is usually done by choosing  $\mu$ so that the integral (\ref{e:RM}) produces results 
in agreement with a known solution obtained with a different method \cite[see, e.g.,][]{Saffman}.
Here we use $\mu=e^{-3/4}$  because this value leads to 
the correct velocity for a ring vortex of uniform vorticity, constant curvature and zero torsion;  
this choice may thus be expected to give a good approximation for a helical vortex
of uniform vorticity, constant curvature and constant torsion.

In the computation of the Rosenhead-Moore integral 
we used a single filament represented by $M$ nodes.
In the computation of the Helmholtz integral we used 100 filaments, each represented by $M$ nodes;
none of these filaments coincides with the vortex centerline so, in contrast to the Biot-Savart case, 
there is no need to smooth out the singularity and, consequently, there is no free tuning parameter
in (\ref{e:Helmholtz}).
The value of $M$ varied between $10^3$ and $10^4$ depending on the helix pitch $L$.

In order to compare the numerical and analytical results, we define 
the relative difference between them as follows
\begin{displaymath}
D_T=\left( \dfrac{ (U_n-U)^2+ (R \Omega_n - R \Omega)^2 } { U^2+(R \Omega)^2} \right)^{1/2}
\end{displaymath}
where quantities without subscript are obtained with formulas (\ref{e:Udim})-(\ref{e:Omdim})
and those with subscript are obtained with either equation (\ref{e:Helmholtz}) or (\ref{e:RM}).
We also find useful to define the relative difference between the linear velocities, $D_U = |U_n - U|/U$,
and the absolute difference between the angular velocities, $\Delta \Omega = |\Omega_n - \Omega|$
(we must use this because $\Omega$ is zero on a curve $\alpha=f(\tau)$, see figure~\ref{f:OmUplane},
and the relative difference would grow indefinitely even if the absolute difference were 
within round-off error).

%
%
%
\begin{figure}
\centering
\includegraphics[width=1.0 \textwidth]{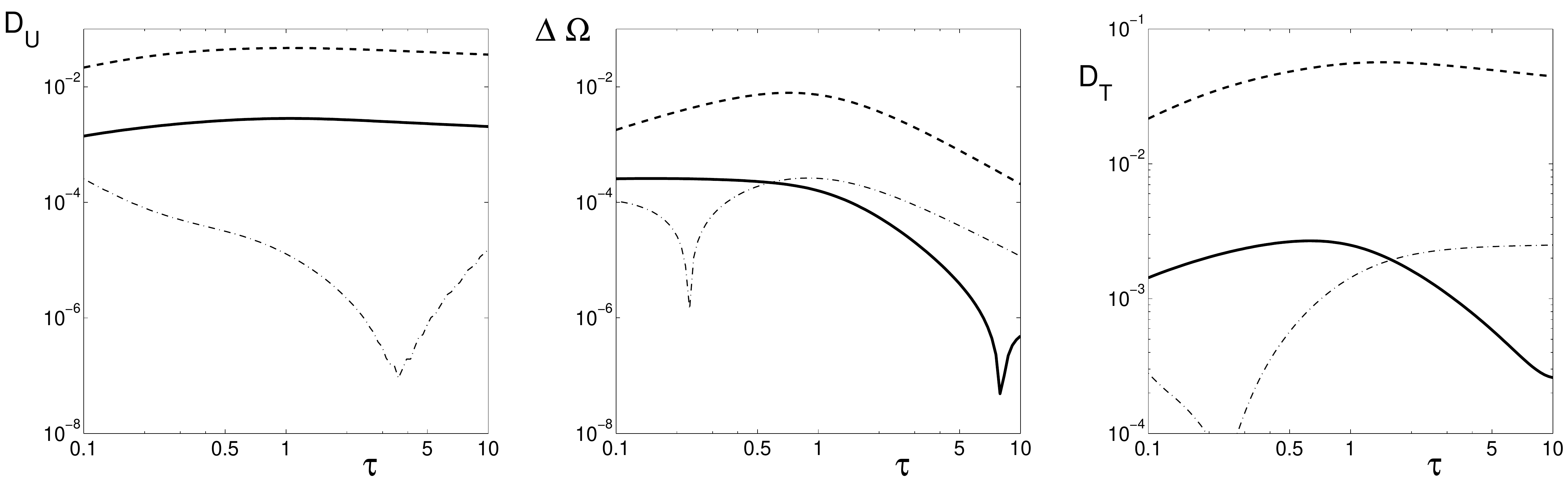}
\caption{
The relative difference in linear velocity $D_U$,
the absolute difference in angular velocity $\Delta \Omega$,
and the relative difference in total velocity $D_T$,
as functions of the vortex pitch ($\tau$),
for a given vortex radius ($\alpha=0.05$).
Results obtained with equations \ref{e:Udim}-\ref{e:Omdim}
are compared with numerical integration of 
equation~\ref{e:Helmholtz}  (continuous line), 
equation~\ref{e:RM} using $\mu=e^{-3/4}$ (dashed line), and
equation~\ref{e:RM} using $\mu=e^{-1}$ (dash-dotted line).
}
\label{f:OmUnum}   
\end{figure}

Figure~\ref{f:OmUnum} shows $D_U$, $D_\Omega$ and $D_T$ as functions
of the vortex pitch ($0.1<\tau<10$) for a given vortex radius ($\alpha=0.05$).
In this case the velocities computed with (\ref{e:Udim})-(\ref{e:Omdim}) and the Helmholtz integral
have a mean relative difference $D_T=0.17 \%$ with a maximum of 0.27 \% at $\tau \approx 0.65$.
In contrast, when the former are compared with Rosenhead-Moore calculations (using $\mu=e^{-3/4}$)
the mean relative difference is 4.6\% with a maximum of 5.7 \% at $\tau \approx 1.4$,
the maximum $D_U$ is 4.7\% at $\tau \approx 1$,
and the maximum $\Delta \Omega$ occurs at $\tau \approx 0.7$.
It is significant that the largest differences
are obtained for vortices with high torsion,
for it must be remembered that $\mu=e^{-3/4}$ is tuned to produce the correct velocity of a vortex of zero torsion.
Therefore we varied the parameter $\mu$ in order to achieve a better agreement and found that
$\mu=e^{-1}$ produces a mean relative difference $D_T=0.13\%$ with a maximum of 0.25\%.

Identical results in $D_U$, $D_\Omega$ and $D_T$ are obtained if the Rosenhead-Moore kernel is replaced by the 
Winckelmans-Leonard kernel \citep{Winckelmans}; i.e., if instead of equation~(\ref{e:RM}) we use
\begin{displaymath}
\boldsymbol{u}(\boldsymbol {x}) = -\dfrac{\Gamma}{4 \pi}  \int 
                     \dfrac{  |\boldsymbol {x}-\boldsymbol {r}(s)|^2+ \frac{5}{2} \mu^2a^2  }
                         {\left( |\boldsymbol {x}-\boldsymbol {r}(s)|^2+\mu^2a^2 \right) ^{5/2}}
                         [\boldsymbol {x}-\boldsymbol {r}(s)] \times d\boldsymbol {s} .
\end{displaymath}
With this kernel it is $\mu=e^{-1/4}$ which leads to the correct ring-vortex velocity
and 6\% differences between analytically and numerically computed velocities of helical vortices,  
whereas $\mu=e^{-1/2}$ brings these differences down to 0.25\%.

\subsection{Comparison with previous results}
\label{s:previous}

\cite{Joukowsky} and \cite{DaRios16} found that, to the order of approximation they used,
the self-induced velocity is entirely in the binormal direction. 
They found the binormal velocity, $u_b$, to be given by
\begin{equation}
\label{e:LIA}
u_b= 
\left \{
\begin{array}{l l} 
\dfrac{ \Gamma}{4 \pi R_c} \log\left(\dfrac{2 R_c }{a}\right) & \text{(Joukowsky 1912)}\\
 & \\
\dfrac{ \Gamma}{4 \pi R_c} \log\left(\dfrac{1}{ a}\right) & \text{(Da Rios, 1916)}  
\end{array}
\right.
\end{equation}
where $R_c=R(1+\tau^2)$ is the radius of curvature. 
Note that the formula of \cite{DaRios16} is a straightforward application of the 
localised induction approximation he introduced ten years earlier \cite[][]{DaRios06}.
We decomposed these velocities in cylindrical components in order to 
compute $U$ and $\Omega$ (see the thin blue lines in figure~\ref{f:OmUprev});
they are in reasonably good agreement with (\ref{e:Udim})--(\ref{e:Omdim}), shown in black thick lines,
for $\tau>2$ only. 

\cite{Levy} and \cite{Widnall} computed the linear velocity $U$ and the angular velocity $\Omega$
of helical vortices of small pitch (in the ranges $0.25<\tau<1.25$ and  $0.1<\tau<1$, respectively).
\cite{Levy} avoided the singularity in the Biot-Savart law by evaluating part of the integral
near, instead of on, the filament;
\cite{Widnall} obtained the self-induced velocity using the cut-off method and
matched asymptotic expansions.
In the range $0.3<\tau<1$ the results of \cite{Widnall} and \cite{Levy}, 
shown in dashed and dash-dotted lines in figure~\ref{f:OmUprev}, 
are in better agreement with equations (\ref{e:Udim})-(\ref{e:Omdim}) than any earlier or later result.

%
%
%
\begin{figure}
\centering
\includegraphics[width=1.0 \textwidth]{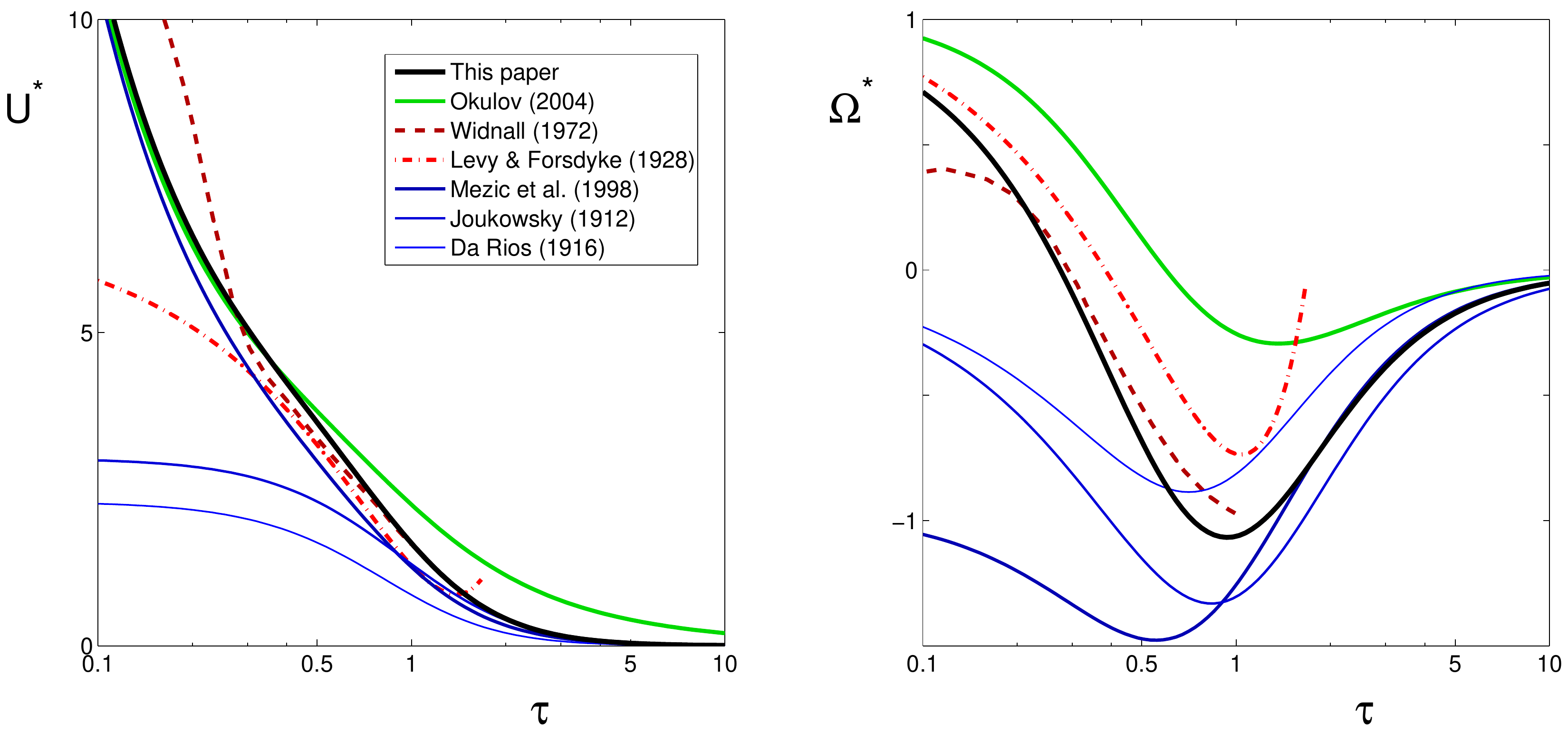}
\caption{
Linear and angular velocities, $U$ and $\Omega$, respectively, 
as functions of the vortex pitch ($\tau$), for a given vortex radius ($\alpha=0.1$).
}
\label{f:OmUprev}   
\end{figure}

\cite{Ricca}, \cite{Boersma} and \cite{Mezic} argued that the tangential component is unimportant
and computed the binormal component only. 
It is true that the former contributes nothing to the time evolution of the Eulerian velocity field but 
it cannot be neglected in the analysis of particle motion (in section~\ref{s:topology}
we will show that neglecting the tangential component, which can reach up to 30~\% of the value of 
the binormal component for relatively thick vortices of small pitch, leads to significant errors in
the determination of the flow topology and the capacity of the vortex to carry fluid).
\cite{Mezic} decomposed the binormal velocity $u_b$ to obtain $U$ and $\Omega$:
the former shows reasonable agreement with (\ref{e:Udim}) for all values of the vortex pitch, 
the latter shows good agreement with (\ref{e:Omdim}) for $\tau>1$ only.

The results of \cite{Okulov} for the binormal component
($u_b$, not shown here) are in good agreement with ours; his results for $U$ and $\Omega$, 
shown in green lines in figure~\ref{f:OmUprev}, 
differ greatly from ours as a consequence of his use of inconsistent assumptions, as explained below. 
In what follows  the prefixes ``O'' and ``R'' indicate
equations in \cite{Okulov} and \cite{Ricca}, respectively.
Equation O4.4, which gives $\Omega$ as a function of $\tau$ and $u_b$,
was obtained using  O2.4 and transformation relations between cylindrical 
and natural components (see R4.1).
Then $u_b$ was computed under the assumption that the vorticity is uniform on the 
vortex cross section (see O4.6 and R4.12). 
A conflict exists because O2.4 implies O2.9, which means
that the vorticity is parallel to helical lines of constant dimensional pitch ($l=L/2\pi$).
Such a vorticity field cannot be uniform on the vortex cross section; thus O4.4 gives an incorrect value of $\Omega$.
The same conflicting assumptions are used by \cite{Okulov2007,Okulov2010}.

%
%
\section{Flow topology}
\label{s:topology}

With an appropriate change of variables it is possible to write the velocity field 
$(u_r,u_{\theta},u_z)$ in terms of a stream function \citep{Hardin}.
If we define $\phi=\theta-z/l$ then $u_\phi=u_\theta-r u_z / l$ and
the velocity field is given by
\begin{eqnarray*}
u_r         &=&\frac{1}{r}\frac{\partial \psi}{\partial\phi}\\
u_{\phi}&=&- \frac{\partial \psi}{\partial r}\\
\end{eqnarray*}
with the stream function defined as follows \citep{Hardin}:
\begin{equation*}
\psi(r,\phi)= 
\left \{
\begin{array}{l l} 
\dfrac{\Gamma (r^2-R^2)}{4\pi l^2}-\dfrac{\Gamma R r}{\pi l^2} S_3(r,\phi)  & \text{if } r<R\\
  & \\
\dfrac{\Gamma}{2\pi}\log \left(\dfrac{R}{r} \right)- \dfrac{\Gamma R r}{\pi l^2} S_4(r,\phi) & \text{if } r>R
\end{array}
\right.
\end{equation*}
where
\begin{eqnarray*}
S_3(r,\phi)&=& \sum_{m=1}^{\infty} K'_m \left( \frac{m R}{l} \right) I'_m \left( \frac{m r}{l}\right) \cos m\phi  \\
S_4(r,\phi)&=& \sum_{m=1}^{\infty} K'_m \left( \frac{m r}{l} \right)  I'_m \left( \frac{m R}{l}\right) \cos m\phi 
\end{eqnarray*}
The stream function gives information about particle motion only if the flow is 
stationary. 
The flow induced by a helical vortex
is time dependent in a reference frame fixed in space,
but it can be made time independent in an infinite number of moving frames.
The most natural choice, and the only one giving information about the capacity of the vortex to carry fluid, 
is a reference frame where the vortex is stationary; that is to say, 
in a frame that translates with linear velocity $U$ and rotates with angular 
velocity $\Omega$. The steady stream function $\Psi$ is obtained with the simple transformation
\begin{equation}
 \Psi=\psi+\frac{1}{2}\left(\Omega-\frac{U}{l}\right) r^2
\end{equation}
%

%
%
%
\begin{figure}
\centering
\includegraphics[width=0.6\textwidth]{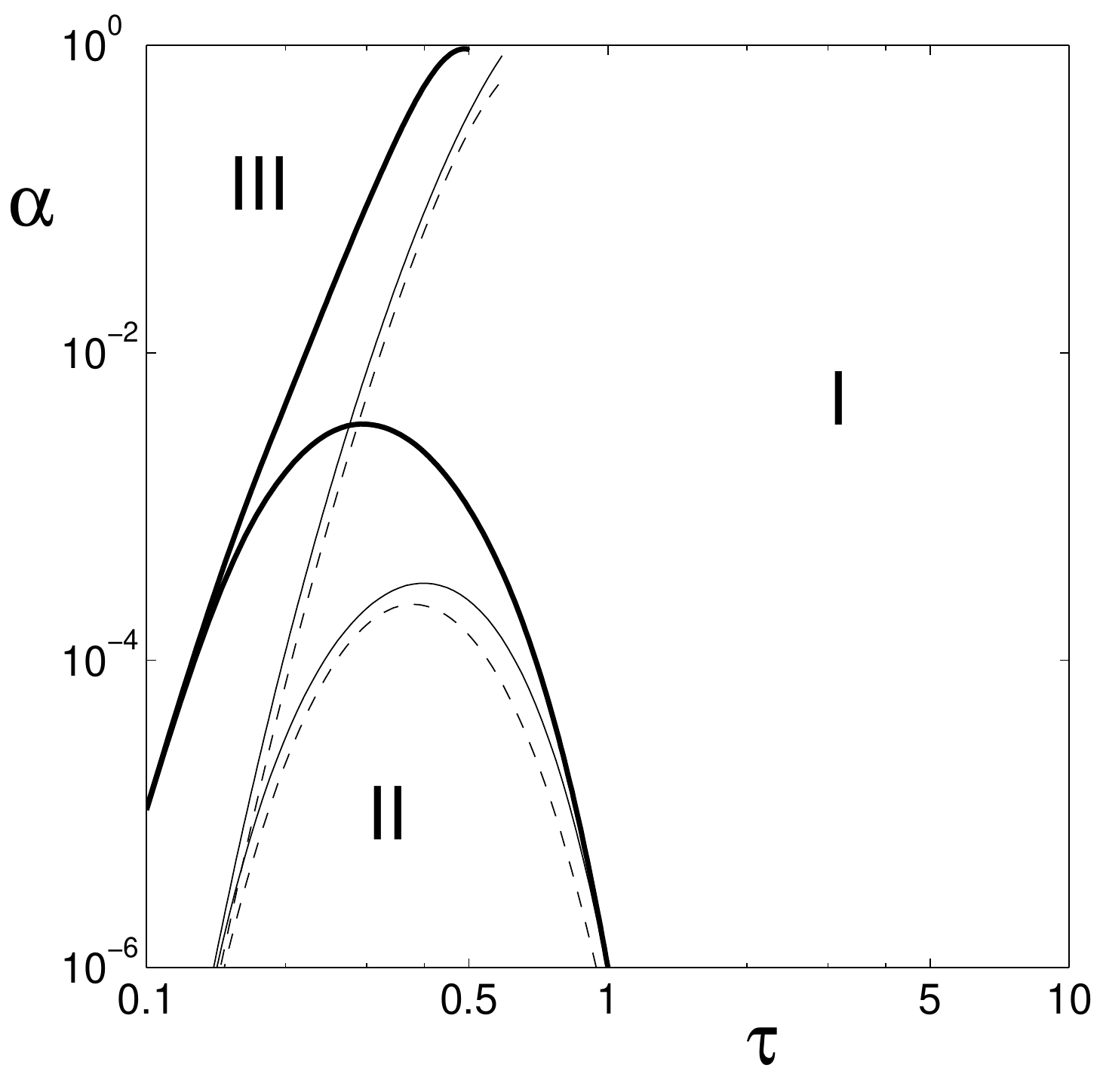}
\caption{
The three flow regimes in the parameter plane $(\tau,\alpha)$: 
(I) large-pitch helices;
(II) thin small-pitch helices; and 
(III) thick small-pitch helices.
The thick lines show the boundaries found in this work,
the thin lines show those found by \cite{Mezic} and
the dashed lines show the results of \cite{Andersen}.
}
\label{f:regimes}   
\end{figure}

Because of the definition $\phi=\theta-z/l$, the curves of constant $\Psi$,  which are streamlines of the helical flow $(u_r,u_\phi)$, 
represent the intersections of the stream surfaces of the three-dimensional flow $(u_r,u_\theta,u_z)$ with the polar plane $z=0$. 
The intersections of the same stream surfaces with the meridional planes $\theta=0,\pi$ can be readily obtained using the same 
definition. 

The flow topology, or phase portrait in the language of dynamical-systems, consist of the set of stagnation points 
plus the streamlines that divide the flow in regions of qualitatively different behaviour. In the co-moving frame 
the vortex, located at $(r,\theta)=(a,0)$,
corresponds to a stagnation point of elliptic type and the symmetries of $\Psi$ indicate that other fixed points, when they 
exist, are located on the line $\theta=0,\pi$. We found these stagnation points using a numerical method 
in order to perform a systematic exploration of the region of the parameter space defined by 
$10^{-6}<\alpha<0.4$ and $0.1<\tau<10$. 

We found three qualitatively different flow topologies (figure~\ref{f:regimes}).
These were first identified by \cite{Mezic} and \cite{Andersen};
they, however, only took into account the binormal component of the vortex motion:
\cite{Mezic} computed $u_b$ using the results of \cite{Ricca}
whereas \cite{Andersen} used equation~(\ref{e:LIA}) with $log(R_c/a)$.
As shown by the blue lines in figure~\ref{f:OmUprev} 
all calculations that do not include the tangential component  
give large errors in $U$ and $\Omega$, 
particularly for small values of the vortex pitch ($\tau<1$). 
Consequently \cite{Mezic} and \cite{Andersen} found regime boundaries that are 
significantly shifted from the ones found here (see the thin lines in figure~\ref{f:regimes}).

%
%
%
%
\begin{figure}
\centering
\includegraphics[width=0.7\textwidth]{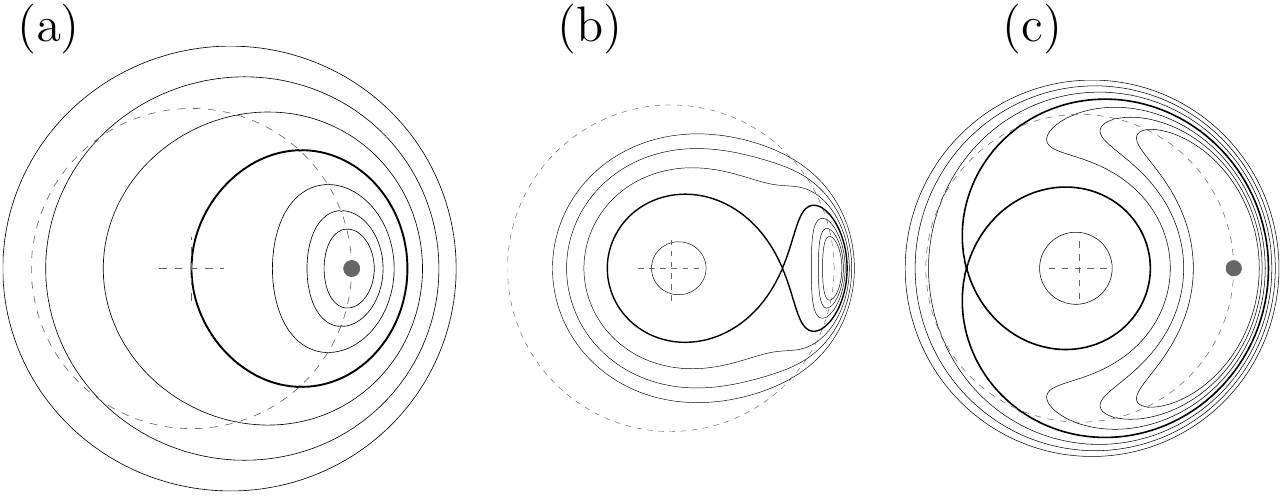}
\caption{
The helical stream function $\Psi$ on the polar plane $(r,\theta)$ for representative cases of each regime: 
(a)  Large-pitch helices ($\tau$=0.8, $\alpha$=0.05) , 
(b) thin small-pitch helices ($\tau$=0.3, $\alpha$=0.0001), and 
(c) thick small-pitch helices ($\tau$=0.2, $\alpha$=0.05).
}
\label{f:psiRTh}   
\end{figure}
%
%
%
\begin{figure}
\centering
\includegraphics[width=0.7\textwidth]{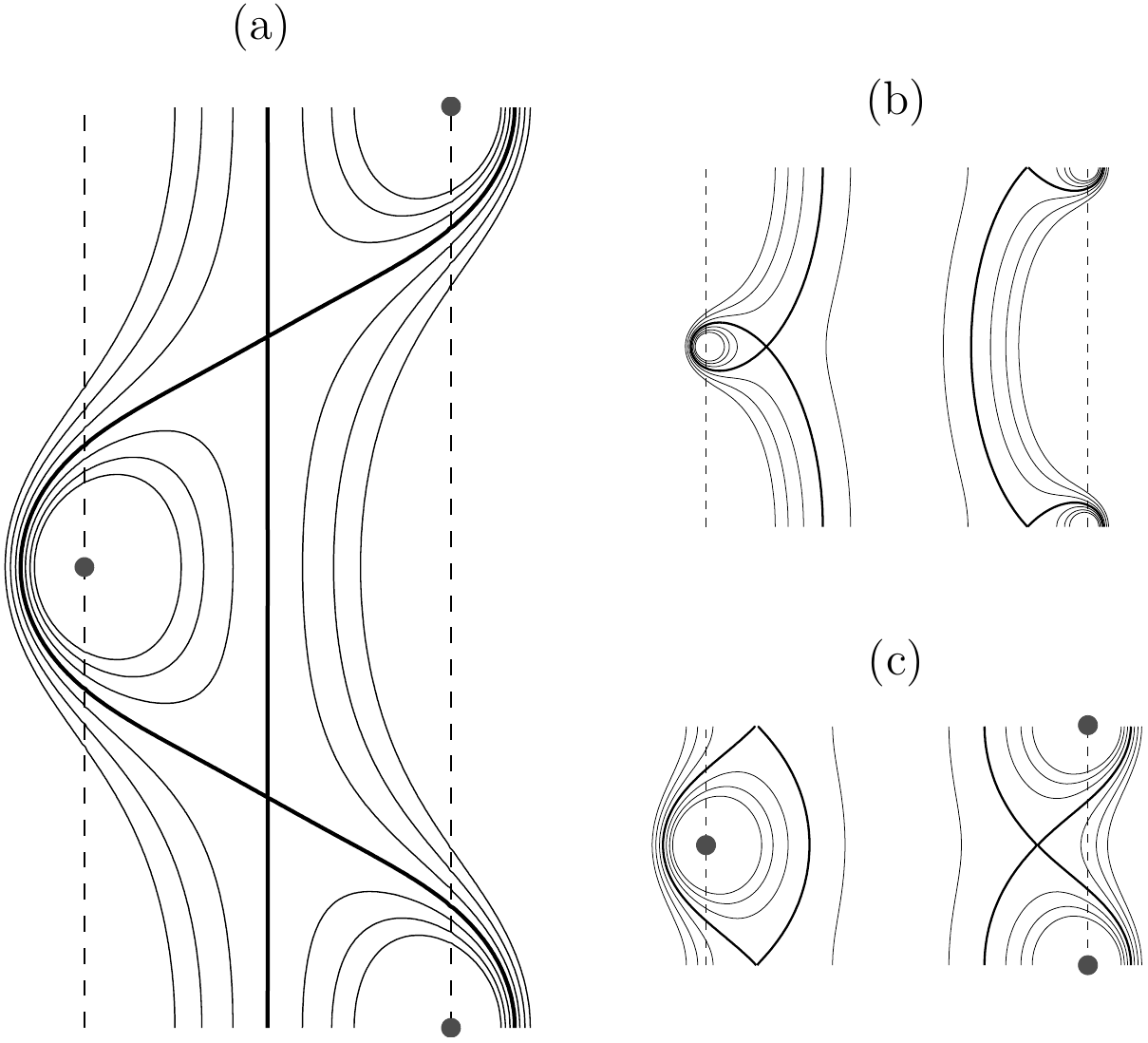}
\caption{
Same as figure~\ref{f:psiRTh} but now on the meridional planes $(r,z)$ corresponding to $\theta=0,\pi$.
}
\label{f:psiRZ}   
\end{figure}

The properties of each flow topology, and the consequences for the vortex's capacity to carry fluid,
are the following:

Regime I (which occupies mostly $\tau>1$ for all values of $\alpha$, see figure~\ref{f:regimes})
is characterised by the existence of only one stagnation point in the plane $r-\theta$, this is of elliptic type and coincides
with the vortex itself (see figure~\ref{f:psiRTh}a). The presence of only one elliptic stagnation point implies that all
particles have, qualitatively, the same behaviour: they all rotate around the vortex.
There is, however, an important quantitative difference: while the particles that are close to the vortex
drift slowly downwards, the particles that are farther from it drift with a speed that rapidly approaches $-U$
(figure~\ref{f:psiRZ}a). In the fixed frame this means that nearby particles follow the vortex in its translation
while distant particles remain close to the horizontal plane where they were initially located.

Regime II (which occupies roughly $\tau<1$ and $\alpha<0.003$, see figure~\ref{f:regimes})
is characterised by the existence of two elliptic stagnation points and a hyperbolic stagnation 
point between them (see figure~\ref{f:psiRTh}b). Now there are three regions with qualitatively different behaviour:
one set of particles rotate  around the vortex and are permanently trapped by it; 
a second set of particles rotate around a line close to the helix axis while drifting downwards;
and the exterior particles rotate around those two sets while drifting downwards (see figure~\ref{f:psiRZ}b).

Regime III (which occupies roughly $\tau<0.4$ and $\alpha>0.0001$, see figure~\ref{f:regimes})
is characterised by the existence of two elliptic stagnation points and a hyperbolic stagnation 
point next to them (see figure~\ref{f:psiRTh}c). There are again three regions with qualitatively different behaviour:
one set of particles rotate anti-clockwise around the vortex and are permanently trapped by it; 
a second set of particles rotate clockwise around a line close to the helix axis while drifting upwards; 
and the exterior particles rotate anti-clockwise around those two sets while drifting downwards (see figure~\ref{f:psiRZ}c).
This is the type of flow \cite{Fitzgerald} speculated about.

\section{Conclusions}
\label{s:conclusions}

We have found new expressions for the linear and angular velocities 
of helical vortices that have, to leading order, uniform vorticity and circular cross-section. 
These  expressions are valid for vortices of any pitch ($\tau$) when their cross-sectional radius ($\alpha$) 
is much smaller than their radius of curvature (i.e., $\alpha \ll 1+\tau^2$).

Numerical computations of the motion of helical vortices using the Helmholtz integral 
and the Rosenhead-Moore approximation to the Biot-Savart law
show that the new formulas give the correct velocities or, at least, very good approximations. 
Indeed, if we assume that $\mu=e^{-3/4}$ in the Rosenhead-Moore kernel leads to the correct velocity
then the best approximations are given by
the new formulas, within 6~\%; \cite{Widnall}, 13.6~\%; and \cite{Levy}, 13.9~\%.
If, on the other hand, we assume that it is $\mu=e^{-1}$ that leads to the correct velocity 
then the best approximations are given by
the new formulas, within 0.5~\%; \cite{Widnall}, 14.1~\%; and \cite{Mezic}, 17.8~\%.
In any case, equations (\ref{e:Udim})-(\ref{e:Omdim}) give the best approximation published so far.

The pitch and radius of the vortex determine its motion and capacity to carry fluid in the following manner:
large-pitch vortices, whether thin or thick, translate slowly and carry with them a large volume of fluid;
thin small-pitch vortices  translate fast and take with them a small volume of fluid; 
and thick small-pitch vortices  translate at intermediate velocities,
take with them a moderate volume of fluid and, more importantly, push fluid forward along the helix axis.

\section*{Acknowledgments}
This research was partially supported by CONACyT (M\'exico)  under grant number 169574.

\end{document}